\documentclass[twocolumn,showpacs]{revtex4} 

\usepackage{graphicx}
\usepackage{dcolumn}
\usepackage{amsmath}

\begin{document}

\title{Lattice distortions, incommensurability, and stripes
in the electronic model for high-T$_c$ cuprates
}

\author{Takashi Yanagisawa, Mitake Miyazaki, Shigeru Koikegami, Soh Koike and
Kunihiko Yamaji 
}

\affiliation{Nanoelectronics Research Institute, National Institute of 
Advanced Industrial Science and Technology (AIST), Central 2, 1-1-1 Umezono, 
Tsukuba, Ibaraki 305-8568, Japan
}

\date{\today}

\begin{abstract}
Striped superconductivity (SC) with lattice distortions is investigated based on 
the three-band Hubbard model for high-T$_c$ cuprates.
A stable inhomogeneous striped state is determined in the low-temperature
tetragonal (LTT) phase
with lattice distortions using a quantum variational Monte Carlo method.  
The ground state has vertical or horizontal hole-rich arrays coexisting with
incommensurate magnetism and SC induced by several percents of lattice deformations.
The SC order parameter oscillates according to the inhomogeneity in the
antiferromagnetic background with its maximums in the hole-rich regions, and 
the SC condensation energy is reduced as the doping rate decreases. 
\end{abstract}


\maketitle

\newpage

\begin{figure}[bp]
\includegraphics[width=\columnwidth]{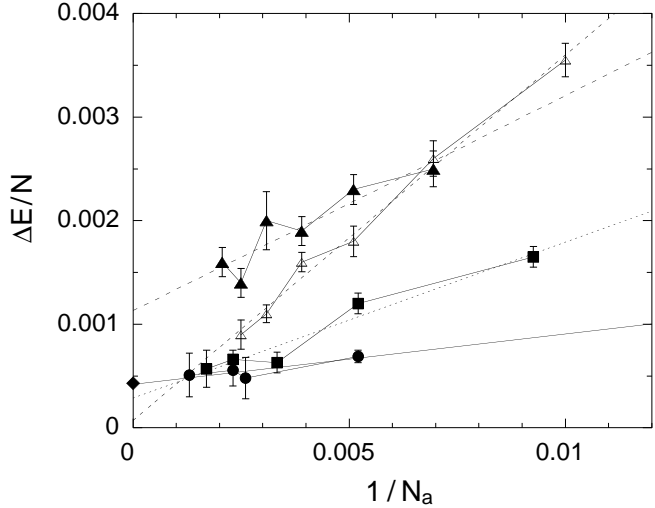}
\caption{
SC condensation energy per site as a function of $1/N_a$ in $t$ units where
$t\approx t_{pd}/3$ and
$N_a$ is the number of atoms.  Squares are for $\delta\approx 0.2$, 
$t_{pp}/t_{pd}=0.4$
and $U_d/t_{pd}=8$ for the three-band model on square lattices.
Circles are at $\delta=1/8$ coexisting with stripes for $t_{pp}/t_{pd}=0.4$ and
$U_d/t_{pd}=8$ on rectangular lattices $32\times 8$, $24\times 6$, $16\times 8$ and
$16\times 4$.
Triangles are for the single-band Hubbard model; $\delta=0.86$ and
$t'=-0.2$ and $U=8$ for solid symbols and $\delta=0.84$ and $t'=-0.15$ for open
symbols (energy unit is $t$).\cite{yam00}
The diamond shows the value indicated from experiments.
}
\label{fig1}
\end{figure} 
 
Over the last decade the oxide
high-T$_c$ superconductors have been investigated intensively.\cite{lt99}
The mechanism of superconductivity (SC) has been extensively studied using
various two-dimensional (2D) models of electronic interactions.
The 2D three-band Hubbard model is the simple and most fundamental model among
such models.  The 2D one-band Hubbard model is regarded as the simplified model
of the three-band model.
Studies of  these models over the last decade indicated that the $d$-wave SC is 
induced from
the electronic repulsive interaction\cite{bic89,pao94,mon94,nak97,yam98,kon01,koig01};
significantly it has been
shown that the SC condensation energy and the magnitude of order parameter
are in reasonable agreement with the experimental results in the optimally
doped case.\cite{yam00,yan00,yan01}

The SC condensation energy obtained by the variational Monte Carlo method (VMC) is
estimated as $E_{cond}\simeq 0.00117t=0.59$meV per site in the optimally doped case
for the single-band Hubbard model in the bulk 
limit.\cite{yam98,yam00}
We must note that $E_{cond}$ is given as $0.17\sim 0.26$meV
by specific heat data\cite{lor93,and98} and $0.26$meV by the critical magnetic field 
value $H^2_c/8\pi$\cite{hao91}.
The agreement of the VMC value with the experimental estimation is quite significant 
and supports the calculations.
The VMC method can be regarded as an approximation to Quantum Monte Carlo
calculations.\cite{yan98,yan99}  
The SC order parameter $\Delta_s$ determined from a minimum of the energy is
of the order of  0.01$\sim$0.015=15meV$\sim$20meV at hole density 
$\delta\sim 0.2$.\cite{yan01}

The interplay between magnetism and superconductivity is suggested
in the underdoped region. 
The reduction of T$_c$ in this region remains unresolved and may be related to 
magnetism.
An existence of incommensurate correlations with modulation vectors given by
$Q_s=(\pi\pm 2\pi\delta,\pi)$ and $Q_c=(\pm 4\pi\delta,0)$ (or
$Q_s=(\pi,\pi\pm 2\pi\delta)$ and $Q_c=(0,\pm 4\pi\delta)$) has been suggested
for the hole-doping rate 
$\delta$.\cite{tra95,suz98,yama98,ara99,wak00,mat00,moo00} 
The linear doping dependence of incommensurability in the underdoped region
supports a striped structure and suggests a relationship between magnetism and
SC.\cite{yama98}
A relationship between the SDW, CDW orders and a crystal structure is also
suggested in intensive studies by the neutron-scattering 
measurements;\cite{tra95,ich00,lee99,kim99} in particular
in the low-temperature tetragonal (LTT) and low-temperature less-orthorhombic (LTLO)
phases, the CDW order is stabilized,\cite{fuj01} while no well 
defined incommensurate CDW peaks were observed for the orthorhombic 
systems.\cite{lee99,kim99}
In the elastic and inelastic neutron scattering experiments with
La$_{2-x}$Sr$_x$CuO$_4$, the incommensurate magnetic scattering spots around
($\pi$,$\pi$) have been observed in the SC phase in the range of
$0.05<x<0.13$.\cite{yama98,kim99,fuj01b}

\begin{figure}[bp]
\includegraphics[width=\columnwidth]{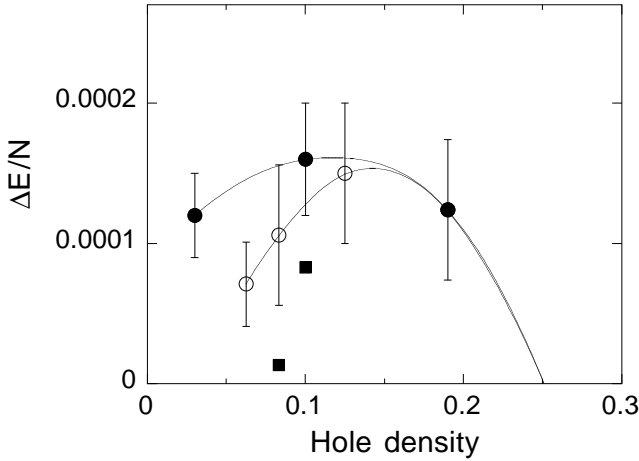}
\caption{
SC condensation energy per site vs the hole density in $t_{pd}$ units, where
the parameters are $t_{pp}=0.4$ and $U_d=8$.
Solid circles and open circles indicate the SC condensation energy for the
uniform SC and striped SC, respectively.
The lines are fitted by parabola.
Squares are obtained for the single-band Hubbard model with the next-nearest
transfer $t'=-0.2$ on $12\times 12$ lattice.\cite{miy02}
}
\label{fig2}
\end{figure}

In this paper an incommensurate striped SC under lattice distortions is
investigated based on the three-band model.
In the description of SC in the underdoped region it is of prime importance to
investigate the effect of inhomogeneity.  In this paper we take into account
all of the inhomogeneity, lattice instability and anisotropic pairing
to clarify the ground state in the underdoped region of high-T$_c$
cuprates.
The Hamiltonian is given by $H=H_{pd}^0+V+H_{el}$ where
\begin{eqnarray}
H_{pd}^0&=& \epsilon_d\sum_{i\sigma}d^{\dag}_{i\sigma}d_{i\sigma}
+ \epsilon_p\sum_{i\sigma}(p^{\dag}_{i+\hat{x}/2,\sigma}p_{i+\hat{x}/2,\sigma}
\nonumber\\
&+& p^{\dag}_{i+\hat{y}/2,\sigma}p_{i+\hat{y}/2,\sigma})
+ t_{pd}\sum_{i\sigma}[d^{\dag}_{i\sigma}(p_{i+\hat{x}/2,\sigma}
+p_{i+\hat{y}/2,\sigma}\nonumber\\
&-& p_{i-\hat{x}/2,\sigma}
- p_{i-\hat{y}/2,\sigma})+h.c.]\nonumber\\
&+& t_{pp}\sum_{i\sigma}(1+v_i)[p^{\dag}_{i+\hat{y}/2,\sigma}p_{i+\hat{x}/2,\sigma}
-p^{\dag}_{i+\hat{y}/2,\sigma}p_{i-\hat{x}/2,\sigma}\nonumber\\
&-&p^{\dag}_{i-\hat{y}/2,\sigma}p_{i+\hat{x}/2,\sigma}
+p^{\dag}_{i-\hat{y}/2,\sigma}p_{i-\hat{x}/2,\sigma} +h.c.].\nonumber\\
&+& t_{pd}\sum_{i\sigma}[u_{i\hat{x}}d^{\dag}_{i\sigma}p_{i+\hat{x}/2,\sigma}
- u_{i,-\hat{x}}d^{\dag}_{i\sigma}p_{i-\hat{x}/2,\sigma}\nonumber\\
&+& u_{i\hat{y}}d^{\dag}_{i\sigma}p_{i+\hat{y}/2,\sigma}
-u_{i,-\hat{y}}d^{\dag}_{i\sigma}p_{i-\hat{y}/2,\sigma}+h.c.],
\end{eqnarray}
\begin{equation}
V= U_d\sum_id^{\dag}_{i\uparrow}d_{i\uparrow}d^{\dag}_{i\downarrow}d_{i\downarrow},
\end{equation} 
where $H_{el}$ denotes the lattice elastic energy given by
$H_{el}=(K_{pd}/2) \sum_{i}(u_{i\hat{x}}^2+u_{i,-\hat{x}}^2+u_{i\hat{y}}^2
+u_{i,-\hat{y}}^2)
+(K_{pp}/2)\sum_i 4v_i^2$ where $K_{pd}$ and $K_{pp}$ denote the elastic
constants.
$\hat{x}$ and $\hat{y}$ represent unit vectors in the $x$- and $y$-direction,
respectively,
$p^{\dag}_{i\pm\hat{x}/2,\sigma}$
and $p_{i\pm\hat{x}/2,\sigma}$ denote the operators for the $p$ electrons at
the site $R_i\pm\hat{x}/2$, and in a similar way $p^{\dag}_{i\pm\hat{y}/2,\sigma}$
and $p_{i\pm\hat{y}/2,\sigma}$ are  defined.
$U_d$ denotes the strength of Coulomb interaction between the $d$ electrons.
$u_{i\hat{\mu}}$ and $v_i$ represent the variations of the transfer energy
$t_{pd}$ and $t_{pp}$, respectively.  
The number of cells which consist of $d$, $p_x$ and $p_y$ orbitals is
denoted as $N$.

\begin{figure}[bp]
\includegraphics[width=\columnwidth]{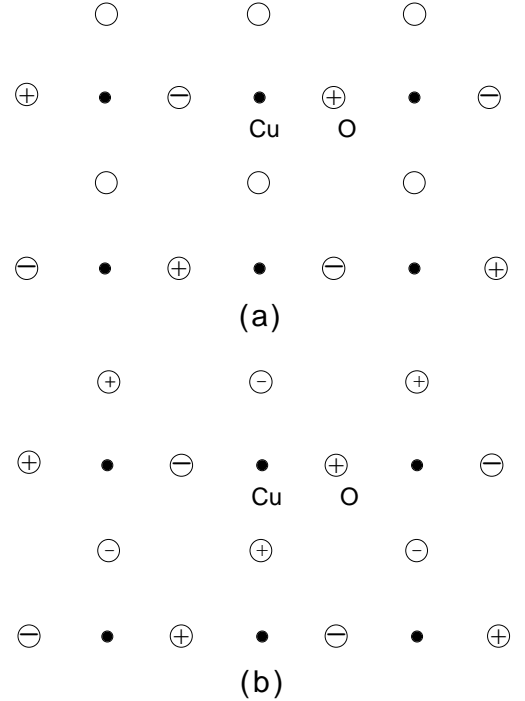}
\caption{
Lattice structures in the LTT phase (a) and LTO phase (b).
The symbol "+" means that the oxygen atoms move upward and instead "-" oxygen
atoms move downward. "O" denote the oxygen atom. 
}
\label{fig3}
\end{figure} 

The wave function with the inhomogeneous spin structure is made from solutions of
the Hartree-Fock Hamiltonian given as
$H_{trial}=
H^0_{pd}+\sum_{i\sigma}[\delta n_{di}-\sigma(-1)^{x_i+y_i}m_i]
d^{\dag}_{i\sigma}d_{i\sigma}$,
where we have variational parameters $\tilde{\epsilon_p}$ and 
$\tilde{\epsilon_d}$ in $H^0_{pd}$. 
In this paper $\delta n_{di}$ and $m_i$ are assumed to have the
form\cite{mac84,gia90,matv00,yan02}:
$\delta n_{di}=-\sum_j \alpha/{\rm cosh}((x_i-x^{str}_j))$,
and
$m_i=\Delta_{inc}\prod_j {\rm tanh}((x_i-x^{str}_j))$,
with parameters $\alpha$ and $\Delta_{inc}$ where
$x^{str}_j$ denote the position of a stripe. 
The inclusion of stripe order parameters considerably improves the ground-state
energy.  In small clusters the deviation of the energy of striped state from
the exact value is within several percents for the Hubbard model.\cite{yan03}

The wave function is constructed 
from the solution of Bogoliubov-de Gennes equation given by
\begin{equation}
\sum_j(H_{ij\uparrow}u^{\lambda}_j+F_{ij}v^{\lambda}_j)= E^{\lambda}u^{\lambda}_i ,
\end{equation}
\begin{equation}
\sum_j(F^*_{ji}u^{\lambda}_j-H_{ji\downarrow}v^{\lambda}_j)= E^{\lambda}v^{\lambda}_i ,\end{equation}
where $(H_{ij\sigma})$ and $(F_{ij})$ are $3N\times 3N$ matrices including the terms
for $d$, $p_x$ and $p_y$ orbitals.
The Bogoliubov operators are written in the form
\begin{equation}
\alpha_{\lambda}= \sum_i (u^{\lambda}_ia_{i\uparrow}+v^{\lambda}_ia^{\dag}_{i\downarrow})~~~(E^{\lambda}>0),
\end{equation}
\begin{equation}
\alpha_{\bar{\lambda}}= \sum_i (u^{\bar{\lambda}}_ia_{i\uparrow}+v^{\bar{\lambda}}_ia^{\dag}_{i\downarrow})~~~(E^{\bar{\lambda}}<0).
\end{equation}
$a_{i\sigma}$ denotes $d_{i\sigma}$, $p_{i+\hat{x}/2\sigma}$ or 
$p_{i+\hat{y}/2\sigma}$corresponding to the components of $u^{\lambda}_i$ and 
$v^{\lambda}_i$.

\begin{figure}[bp]
\includegraphics[width=\columnwidth]{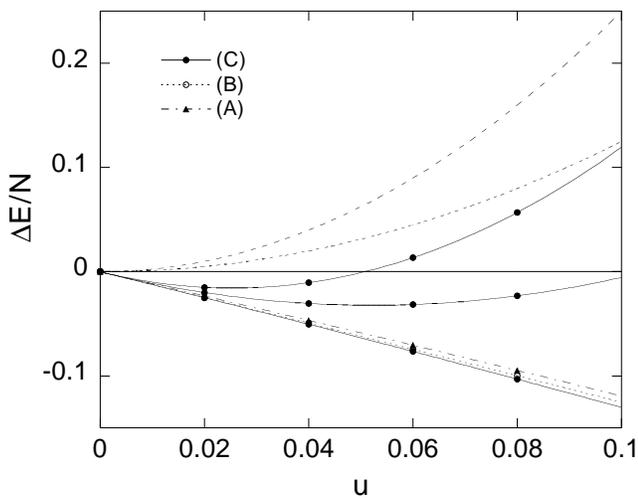}
\caption{
Energy gain $\Delta E=E(u=0)-E(u)$ ($u=\delta t_{pd}/t_{pd}$) per site as a 
function of transfer deformation $u$ in $t_{pd}$ units. 
The parameters are $t_{pp}=0.4$ and $U_d=8$ for $16\times 4$ lattice.
The hole-rich stripes are in the $y$-direction.  The energy gains for (A) 
$u_{i\hat{y}}=0$
 (triangles), (B) $u_{i\hat{x}}=0$,
(open circles) and (C) (solid circles) are shown. 
The elastic energy $Ku^2/2$ is shown by the dashed line (for $K=5$ and $K=10$).
The summations of 
$\Delta E=E(u=0)-E(u)$ and the
elastic energy per site are also shown for the case (C).
The Monte Carlo statistical errors are smaller than
the size of symbols.
}
\label{fig4}
\end{figure}

\begin{figure}[bp]
\includegraphics[width=\columnwidth]{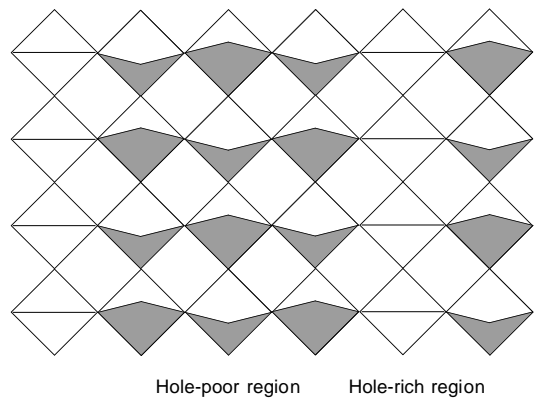}
\caption{
Schematic structure of lattice distortions and stripes 
where the hole-rich arrays are perpendicular to the tilting axis.
We call this state the mixed LTT-HTT phase.
The shaded square represents distorted CuO unit cell. 
}
\label{fig5}
\end{figure}

Then the wave function is written as\cite{him02,yan02b,miy02} 
\begin{eqnarray}
\psi&=& P_G P_{N_e}\prod_{\lambda}\alpha_{\lambda}\alpha^{\dag}_{\bar{\lambda}}|0\rangle\nonumber\\
&\propto& P_G\{\sum_{ij}(U^{-1}V)_{ij}a^{\dag}_{i\uparrow}a^{\dag}_{j\downarrow}\}^{N_e/2}|0\rangle.
\end{eqnarray}
$U$ and $V$ are matrices defined by
$(V)_{\lambda j}=v^{\lambda}_j$ and $(U)_{\lambda j}=u^{\lambda}_j$.
$P_G$ is the Gutzwiller operator.
The spin modulation potential is contained in $(H_{ij\sigma})$ and the SC order
parameters $\Delta_{ij}$ are included in $(F_{ij})$.  We assume the following
spatial variation for the SC order parameters in the $d$-electron part:
$\Delta_{i,i+\hat{x}}= \Delta_s {\rm cos}(Q_x(x_i+\hat{x}/2))$,
$\Delta_{i,i+\hat{y}}= -\Delta_s {\rm cos}(Q_x x_i)$,
where $Q_x=2\pi\delta$ ($\delta$ is the hole density).
The SC order parameter oscillates according to the spin and charge distributions
so that the amplitude has a maximum in the hole-rich region and is suppressed in
the hole-poor region.  
The energy-expectation value is calculated using the Monte Carlo algorithm\cite{yam98,
yan98,yan01}:
$\langle O\rangle = \langle\psi|O|\psi\rangle/\langle\psi|\psi\rangle$.

Here we show the results in the case without lattice distortions.
It has been shown that the striped state is more stable than the uniform SDW
state for small hole doping.\cite{yan02}
In Fig.1 the size dependence of 
the SC condensation energy is shown for the uniform SC in the overdoped
region and the striped SC in the underdoped region with the results obtained for the
one-band Hubbard model for comparison.
The parameters are $t_{pp}=0.4$ and $U_d=8$ in $t_{pd}$ units.
The squares in Fig.1 indicate the SC condensation energy of pure $d$-wave state
at $\delta\approx 0.2$, while the circles are for SC coexisting with stripes at
$\delta=1/8$ for $Q_x=\pi/4$ evaluated on rectangular lattices $32\times 8$,
$24\times 6$, $16\times 8$ and $16\times 4$.
In both cases the energy obtained through an extrapolation is of the same order as 
experimental values.
\begin{equation}
E_{cond} \approx 0.00014t_{pd} \approx 0.2{\rm meV},
\end{equation}
where we have assigned $t_{pd}\approx 1.5$eV.\cite{esk89}
The data in Fig.2 show the SC condensation energy as a function of the hole
density.
The SC condensation energy per site for the striped SC is reduced as the hole density 
decreases, while
that for pure $d$-wave SC remains finite even near half-filling. 
This suggests that an origin of the decrease of T$_c$ in the underdoped region lies 
in the reduction of hole-rich domain where the SC order parameter has finite amplitude.

Now let us consider the effect of lattice distortion on stripes.
In the LTT phase stabilized at low temperatures near 
1/8-hole filling,\cite{suz89} 
the distortions of the 
CuO square occur in the manner shown in Fig.3.
The LTT phase has a 'tilting axis' on which the copper and oxygen atoms never move
even in the distorted state.\cite{bia96} 
The vertical or horizontal
stripes can coexist with the lattice distortions in the LTT phase, being parallel
or perpendicular to the tilting axis.

The structural transition from low-temperature orthorhombic (LTO) to LTT phases
occurs in LaBaSrCuO and LaNdSrCuO systems.\cite{ich00}
It is not clear a priori what structure is stabilized due to the lattice deformation.
We consider the following cases
assuming that the stripes are in the $y$-direction:\\
\\
(A) $u_{i\hat{x}}=u, u_{i\hat{y}}=0, v_i=0$,
\\
(B) $u_{i\hat{x}}=0, u_{i\hat{y}}=u, v_i=0$,
\\
(C) $u_{i\hat{x}}=u, u_{i\hat{y}}=u{\rm cos}(2Q_xx_i),
v_i=u{\rm cos}(2Q_x x_i)$,
\\
\\
where $Q_x=2\pi\delta$ and $u$ is the amplitude of deformation: 
$u_{i\hat{\mu}}=u_{i,-\hat{\mu}}$.
$u=0$ corresponds to the LTO sturcture, and the anisotropy in $t_{pd}$ indicates a
transition to the LTT phase. 
$t_{pd}$ increases along the tilt axis compared to the LTO phase.
The case (C) corresponds to the structure of mixed LTT-HTT phase.
The energy gain per site defined as $\Delta E/N=(E(u=0)-E(u))/N$ is presented 
in Fig.4 as a function of $u$ in $t_{pd}$ units.
The energy in the case (B) is lower than that in the case (A) indicating that
the stripes are parallel to the tilting axis under the rigid LTT structure.
The energy in the cse (B) is lower than that in the case (A) indicating that
the stripes are parallel to the tilting axis under the rigid LTT structure.
We simply assume the same elastic energy cost for these types of rotations.
The cost of energy due to lattice distortions is assumed to be given by
$(K/2)u^2$ for the constant $K$.
$K$ is estimated as follows.
According to Harrison's rule,\cite{har80} $t_{pd}$ is expected to vary as $d^{-n}$ with
$n\approx 7/2$, $d$ being the Cu-O bond length.
Since $\delta t_{pd}/t_{pd}=-n\delta d/d$, the elastic energy is estimated as
\begin{equation}
E_{el}= \frac{1}{2}C (2d)^3 (\frac{\delta d}{d})^2 = \frac{1}{2}C(2d)^3\frac{1}{n^2}u^2
\equiv \frac{K}{2}u^2.
\end{equation}
The constant $C$ is estimated as $C\approx 1.7\times 10^{12}$ dyn/$cm^2$= 
1.7 eV/$\AA^3$.\cite{mig90} Since $d\approx 2\AA$,  $K$ is of the order of 10eV: 
$K\approx 8.9$eV.
We point out that $E_{el}$ has possibly a linear term in $u$.  The $\theta^2$ term,
if present in $E_{el}$, is proportional to $u$ since 
$u\sim 1-{\rm cos}(\theta)\sim \theta^2/2$ where $\theta$ is the tilt angle.  
The presence of linear term may lead
to a first order transition.
As shown in Fig.4 the striped state is more stabilized in the LTT phase.
We show schematically the stable striped state in the LTT phase in Fig.5 
obtained
from our VMC evaluations, where the shaded square represents the tilted CuO unit
cell rotating around the tilting axis.
The LTT-HTT state is more stabilized due to the kinetic energy gain coming from
the softening of tilt angles.

In this paper we have investigated the inhomogeneous ground state with the lattice
distortions based on the three-band model of high-T$_c$ cuprates using the variational
Monte Carlo method.
The SC condensation energy decreases as
the doping ratio decreases, which is due to the reduction of SC domain in the
hole-rich region.
The stable striped state has hole-rich arrays being perpendicular to the tilting axis
of the lattice distortions in the LTT phase as shown in Fig.5, which can be regarded
as the LTT-HTT mixed phase.
We thank H. Oyanagi for valuable discussions.

\end{document}